\documentclass[pra,twocolumn,showkeywords,longbibliography,superscriptaddress,10pt]{revtex4-1}
\usepackage{graphicx}
\usepackage{dcolumn}
\usepackage{color}
\usepackage{times}
\usepackage{bm}
\usepackage{amssymb}
\usepackage{amsmath}
\usepackage{epsfig}
\usepackage{epstopdf}
\usepackage{dsfont}
\usepackage{subfigure}
\usepackage{tikz}
\usepackage[colorlinks=true, citecolor=blue, linkcolor=blue, urlcolor=blue]{hyperref}
\usepackage[mathscr]{euscript}
\makeatletter
\renewcommand{\maketag@@@}[1]{\hbox{\m@th\normalsize\normalfont#1}}
\makeatother
\begin{document}
\renewcommand{\thefootnote}{\fnsymbol {footnote}}

\title{Quantum battery in the Heisenberg spin chain models with Dzyaloshinskii-Moriya interaction}

\author{Xiang-Long Zhang}
\affiliation{School of Physics \& Optoelectronic Engineering, Anhui University, Hefei
230601,  People's Republic of China}

\author{Xue-Ke Song}
\affiliation{School of Physics \& Optoelectronic Engineering, Anhui University, Hefei 230601,  People's Republic of China}
%

\author{Dong Wang} \email{dwang@ahu.edu.cn}
\affiliation{School of Physics \& Optoelectronic Engineering, Anhui University, Hefei
230601,  People's Republic of China}


\date{\today}

\begin{abstract}
{Quantum battery (QB) is an energy storage and  extraction device conforming to the principles of quantum mechanics. In this study, we consider the characteristics of QBs for the Heisenberg spin chain models in the absence and presence of Dzyaloshinskii-Moriya (DM) interaction. Our results show that the DM interaction can enhance the ergotropy and power of QBs, which shows the collective charging can outperform parallel charging regarding QB's performance. Besides, it turns out that first-order coherence is a crucial quantum resource during charging, while quantum steering between the cells is not conducive to the energy storage of QBs. Our investigations offer insight into the properties of QBs with Heisenberg spin chain models with DM interaction and facilitate us to acquire the performance in the framework of realistic quantum batteries.}

\end{abstract}

\maketitle

\section{Introduction}
The rapid development of the quantum information field has brought great changes to the world, and some devices, based on the quantum properties of matter, have shown great advantages compared with traditional devices.
Among them, QBs have attracted wide attention because of their unique quantum advantages \cite{z1,z2,z3,z4,z5,z6,z7,z8,z9}. Alicki and Fannes conducted pioneering work on the proposal of QBs (QBs) as well as the study of quantum effects to improve their performance \cite{z10}.
Like traditional electrochemical cells, QBs are used as temporary energy storage devices; they have a limited energy capacity and power density and may also dissipate due to unavoidably interacting with the
environment \cite{z11,z12,z13,z14}. Conversely, QBs can be charged (or consumed) quantum operations that create a coherent superposition between different states \cite{z15}, and interestingly, the existence of quantum effects allows QBs to have high charging speeds that far exceed those of conventional batteries.

{Typically, QBs have often been defined as a set of $N$ identical and independent subsystems, upon which a temporary charging field acts to either extract or deposit work.
Alicki and Fannes, in particular, demonstrated that the global entangling operations allows for the extraction of more work from a QB compared to local
operations} \cite{z10}. This was further observed by Hovhannisyan et al. \cite{z16}, who found that a series of ${N}$ global entangling operations can extract
the maximum work without creating any entanglement in the QB. {In this theoretical framework, the quantum battery is modeled as a one-dimensional
Heisenberg spin chain consisting of $N$ spins, inherently facilitating interactions among the spins and offers the entanglement} \cite{z17}. Later, two types of charging
schemes, "parallel" and "collective" schemes were proposed \cite{z18,z19}. Some conclusions have emerged about collective and parallel charging methods for QBs.
{For the QB's charging process, a collective charging scheme offers a ``quantum advantage," that is, when ${N} {\ge2 } $ , the charging power of the QB outperforms that of a parallel
charging scheme }\cite{z20,z21,z22,z23,z24,z25,z26,z27,z28,z29,z30,z31,z32}.
{Campaioli et al. demonstrated that a battery pack consisting of $N$ highly mixed qubits can charge collectively $N$ times faster than  each qubit was charged separately} \cite{z33}.

The spin system is regarded as the most popular one for constructing QBs. To date, much attention has been paid for exploiting QBs with spin systems. For example, the spin-spin interactions can enhance charging power compared to noninteracting scenarios in the XXZ Heisenberg chain  \cite{z34}. Dou et al. proposed a proposal of QB in the presence of cavity Heisenberg-spin-chain with long-range interactions, which has significance for the performance improvement of QBs \cite{z25}.
{Shi et al. revealed that quantum coherence in the battery or entanglement between the battery and charger is essential for producing nonzero extractable work during the charging process} \cite{z35}. There is an important term in spin systems, say Dzyaloshinsky-Moriya ({DM}) interaction \cite{z36,z37}, which is deemed as the anisotropic superexchange interaction, and significantly reflects the properties of the system of interest. In this sense, it is fundamentally required and meaningful to unveil how the DM interaction influences the performance of QBs in practice. Motivated by this, our work readily contributes to addressing this issue.





The remainder of this paper is structured as follows. In Sec. II, we describe the charging models of QBs and introduce the measure to quantify QB's performance.
In Sec. III, we examine the influence of quantum correlation in the evolution of the Ising model, {XXZ} model, and {XYZ} model and the influence of the variation of {DM} interaction strength on them.
Finally, a summary is provided in Sec. IV.

\section{charging model}

\begin{figure}[t]
\centering
\subfigure{\includegraphics[width=7cm]{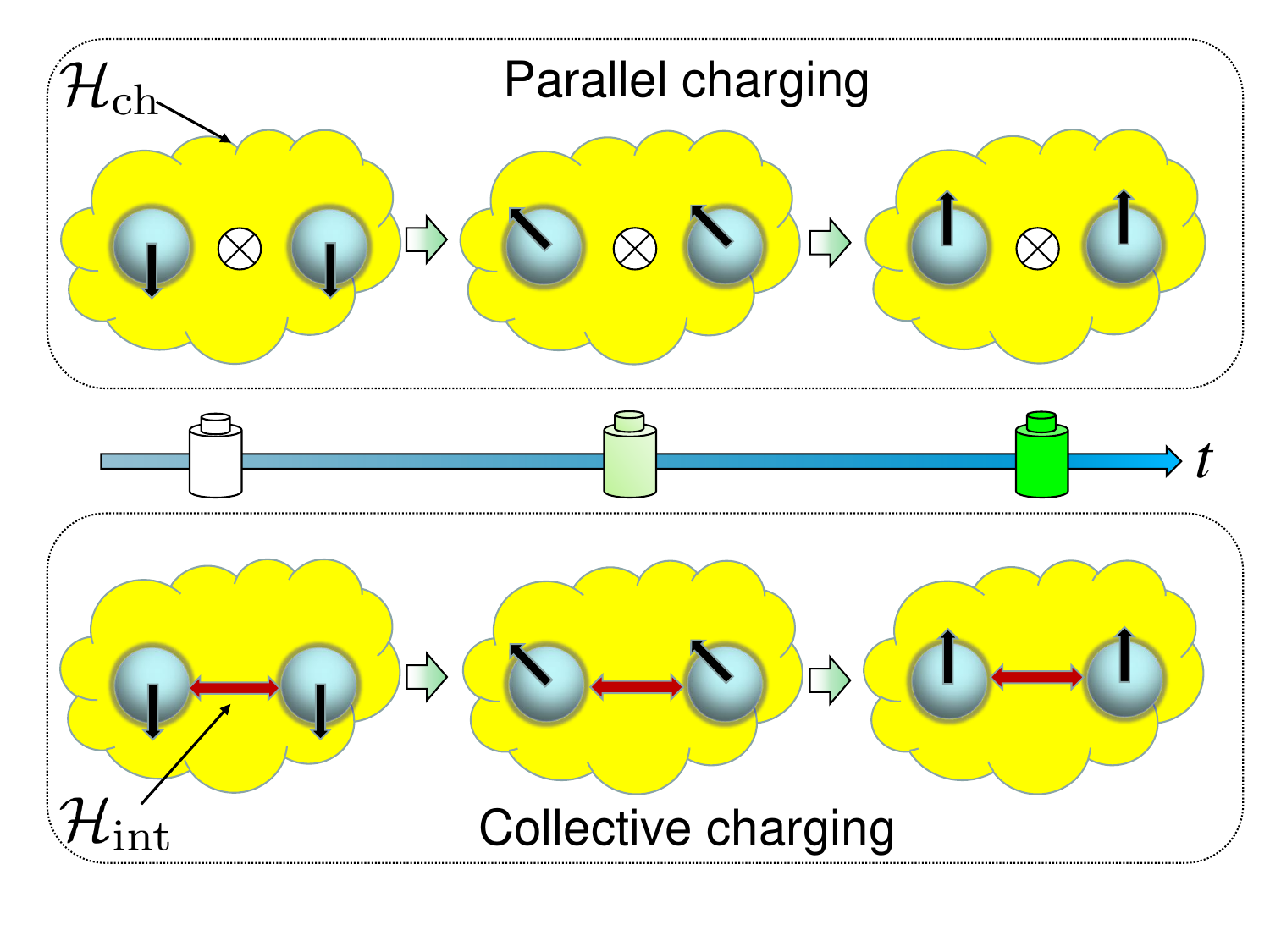}}
\caption{Diagram of two-cell QB, the spin system is charged by parallel and collective charging, respectively. The battery is initially in the ground state (spin-down), and it will be in the excited state (spin-up) for the maximal energy of QB as time $t$ goes by. A local electric field acts on the cells, and the cells interact with each other along the collective charge. When collectively charged, the system can evolve through entangled states.}
\label{f1}
\end{figure}

We consider a two-qubit {QB} model consisting of two coupled two-level systems. The battery, meanwhile, is charged by a local field coupling each cell individually, as shown in Fig. \ref{f1}. Without loss of generality \cite{z38},
the driving Hamiltonian of the {QB} system can be written as (hereafter, we set $\hbar=1$)
\begin{align}
{\cal H}  &= {\cal H}_{\mathrm{ch}} + {\cal H}_{\mathrm{int}}, \\
{\cal H}_{\mathrm{ch}} &= \hbar \Omega\sum_{n=1}^{2} \hat{\sigma}_{n}^{x} ,
\end{align}
where ${\hat{\sigma} _{n}^{x} }$ is the Pauli X-matrix acting on the ${n}$-th spin, and ${\Omega}$ is the strength of the charged magnetic field. The second item $ {\cal H}_{\mathrm{int}}$ denotes the interaction one given by the Heisenberg Hamiltonian with DM interaction. The Hamiltonian of the
Heisenberg chain model with DM interaction in the $z$-direction of 2 sites is
\begin{align}
\begin{split}
{\cal H}_{\mathrm{int}}  =&J\hbar \left[(1+\gamma )\hat{\sigma} _{1}^{x} \hat{\sigma} _{2}^{x}+(1-\gamma )\hat{\sigma} _{1}^{y}\hat{\sigma} _{2}^{y}+\Delta \hat{\sigma} _{1}^{z}\hat{\sigma} _{2}^{z}\right] \\
&+D(\hat{\sigma} _{1}^{x}\hat{\sigma} _{2}^{y}-\hat{\sigma} _{1}^{y}\hat{\sigma} _{2}^{x}) ,
\label{Eq.3}
\end{split}
\end{align}
where ${J}$ is the nearest exchange coupling constant, ${\gamma }$ and ${\Delta}$ are the dimensionless parameters related to chain anisotropy.
${D}$ is the strength of {DM} interaction in the ${z}$-direction, and ${\hat{\sigma}_{i}^{\alpha } } $ (${{\alpha = x,y}} $ ) are Pauli operators of the ${i}$-th site. If $\gamma=1$ and $D=\Delta=0$ are held, the current model is converted to the simple Ising model. If $\gamma=D=0$ and $\Delta \neq 0$, the model becomes the so-called XXZ model. When $\Delta \neq 0$ and $\gamma \neq 0$, the battery is converted into the XYZ model.
For QBs, the free Hamiltonian ${{\cal H}_{0}}$ considered here as ${{\cal H}_0=\hbar w_0 {\textstyle \sum_{n=1}^{2}}\hat{\sigma} _{n}^{z}  }$ plays a crucial role in the ability of storing energy, with identical Larmor frequency ${w_0}$ for both qubits. Here, the ground and excited states of a single spin are represented by ${|\downarrow  \rangle} $ and ${|\uparrow \rangle}$, respectively. Thus, the state of the fully charged battery is ${|\uparrow \uparrow \rangle} $ with energy ${2\hbar w_0 } $, and the empty state is $ |\psi_0\rangle $= ${|\downarrow\downarrow\rangle} $ with energy ${-2\hbar w_0 } $.
The charging process can be described by a unitary operator ${\cal U}$, namely,
\begin{align}
{\cal U}= e^{-i{\cal H}t}
\end{align}
The system dynamics obey the following equation:
\begin{align}
|\psi(t)\rangle = {\cal U}|\psi(0)\rangle.
\label{Eq.5}
\end{align}
The work extracted from a quantum battery is well defined as the ergotropy \cite{z39}, {and here we introduce the concept of a passive state, mentioning the state in which no works can be extracted from them by unitary transformations.} However, for pure states, the lowest energy state of the system is its ground state; therefore, the passive state can be well defined as its ground state \cite{z40}. Here, the ergotropy can be determined by the difference between the final and ground state energies. In addition to the ergotropy, the power is another important indicator to evaluate the QBs' performance. Specifically, the ergotropy and power of the battery can be given by the following formula:
\begin{align}
&\zeta =\mathrm{Tr}(\hat{\rho} {{\cal{H}}_{0}}) -\max   \mathrm{Tr} \mathrm({\cal{U}}\hat{\rho}{\cal{U^{\dagger} }} {{\cal{H}}_{0}}),\\
&P\left(t\right)  = \frac{ {\zeta} }{\mathit{t}},
\end{align}
respectively. First, we study the case where there is no interaction between the two batteries $(J=0)$, and each battery evolves independently driven by the charging field. From this, we can deduce that the ergotropy of the battery during the parallel charging is
\begin{align}
\zeta _\shortparallel (t) & = \zeta_{\max}\sin^{2} (\Omega t).
\end{align}
From the above equation, one can obtain the minimal time $t_{\min}=\pi/(2\Omega)$, which corresponds to the maximal ergotropy of QB in parallel charging.

\section{Steering, first-order coherence and charging power}

To study the quantumness of the two-cell {QB}, we herein detailedly examine quantum steering \cite{z41,z42,z43,z44,z45} and the coherence of the system state.
{Conceptually, quantum steering is logically distinct from entanglement and Bell nonlocality, which represents a strict subset of entangled states
and a strict superset of Bell nonlocal states, and} quantum steering describes the ability of
a local measurement of an entangled particle acting on one of the particles to affect the state of the other particle non-locally \cite{z46}.
For a two-qubit system, it can be expressed as \cite{z47}
  \begin{align}
  \frac{1}{\sqrt{3} } \left | \sum_{i=1}^{3}\mathrm{Tr}(A_i\otimes B_i \hat{\rho  }_{AB} ) \right | \le 1,
  \end{align}
where ${\mathit{A_{i}}=\mathbf{a_{\mathit{i}}\cdot {\mathbf{\hat{\sigma} } } } } $ and ${B_{\mathit{i}}=\mathbf{b_{\mathit{i}}\cdot {\mathbf{\hat{\sigma} } }  } } $ with $i\in \{1,\mathrm{2} ,\mathrm{3} \}$, are Hermitian operators acting on qubits ${A} $ and ${B} $, respectively. Here,
  ${\mathbf{a_\mathit{i} ,b_\mathit{i} } \in \mathbb{R^{\mathrm{3} } } }$ are two unit vectors and ${\mathbf{\hat{\sigma} =\mathrm{(} \mathrm{\hat{\sigma} _1,\hat{\sigma} _2,\hat{\sigma} _3)} } } $  denotes the Pauli matrices. Any violation of inequality implies
  that ${ \hat{\rho  } _{AB} }$ is steerable, and the maximal violation can be described as \cite{z48}
    \begin{align}
    \ S_{AB}=\underset{\left \{ A_{\mathit{i}},B_{\mathit{i}} \right \} }{\max} \frac{1}{\sqrt{3} } \left | \sum_{i=1}^{3}\mathrm{Tr}(A_i\otimes B_i \hat{\rho  }_{AB} ) \right |.
    \end{align}

{In addition, coherence results from the superposition of quantum states and can be used to assess the interference capability of interaction
fields. In optical systems, first-order coherence is a commonly used measure of coherence}. It is crucial to introduce the concept of
first-order coherence, which plays a significant role in understanding quantum correlations \cite{z49,z50,z51,z52}. For the single-qubit state
$ \hat{\rho  }_{k} $, its first-order coherence can be quantified in terms of its purity \cite{z53}, i.e.,
\begin{align}
Q\left(\hat{\rho  } _{k}\right) = \sqrt{2 \mathrm{Tr}\left( \hat{\rho  }^{2}_{k}\right)-1}.
\end{align}
Consider a two-qubit state ${ \hat{\rho  } }$, and its corresponding reduced density matrices ${ \hat{\rho  }_A=\mathrm{Tr}_B( \hat{\rho  } ) } $ and ${ \hat{\rho  }_B=\mathrm{Tr}_A( \hat{\rho  } ) } $ are the
reduced density matrices. The first-order coherence for the state ${ \hat{\rho  } }$  can be determined using

\begin{align}
Q( \hat{\rho} )=\sqrt{ \frac{Q^2{( \hat{\rho  } _A)}+Q^2{( \hat{\rho  } _B)}}{2} }.
\end{align}
In all, both the coherence and steerability can perfectly reflect the quantum characteristics of the system of interest.

\subsection{{Ising model with DM interaction}}
First, we start with the simplest Ising model $({\rm i.e.,}\ \gamma =1,\ \Delta =0)$ and explore the effect of {DM} interactions on the QB's performance. It is necessary to determine a suitable time window to facilitate our research.
As mentioned before, the minimum time for a quantum battery to reach its maximum ergotropy is ${t_{\min}}$. Therefore, to compare the two charging models, parallel and collective charging,
we consider the time window of system evolution as ${\mathbf{ \tau} }  =\left [ 0,t_{\min} \right ] $. Notably, the ergotropy and power of QBs are regarded as two crucial aspects of evaluating the performance of a QB.
Besides, since ${t_{\min}} $ is the minimum charge time for parallel charging if one achieves the maximum charge in a smaller period of time than ${t_{\min}} $, we then say the quantumness of the system, i.e., quantum correlation, plays a positive role in charging, which manifests the quantum advantage.

\begin{figure}[htbp]
\centering
\subfigure{\includegraphics[width=0.5\linewidth]{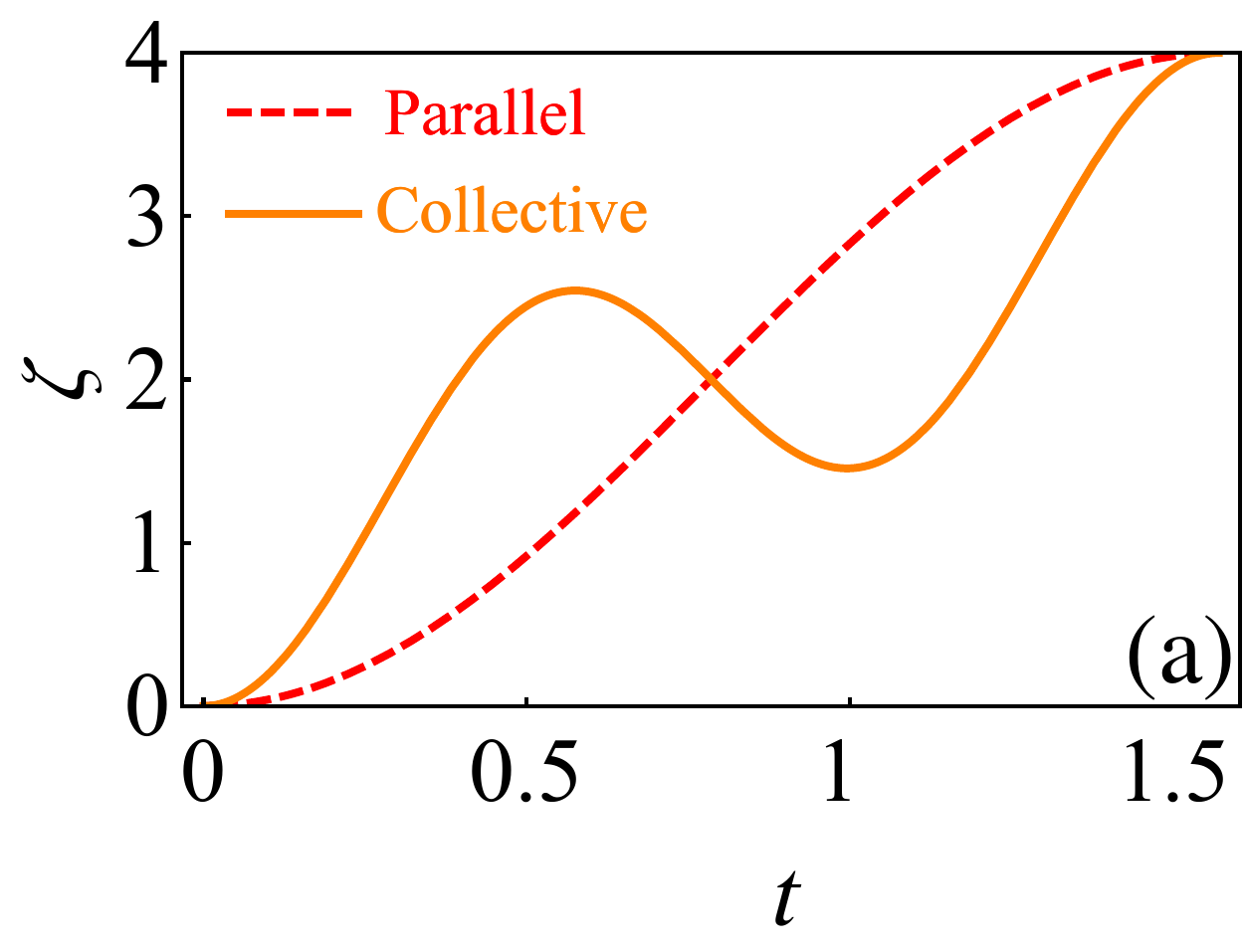}}\subfigure{\includegraphics[width=0.5\linewidth]{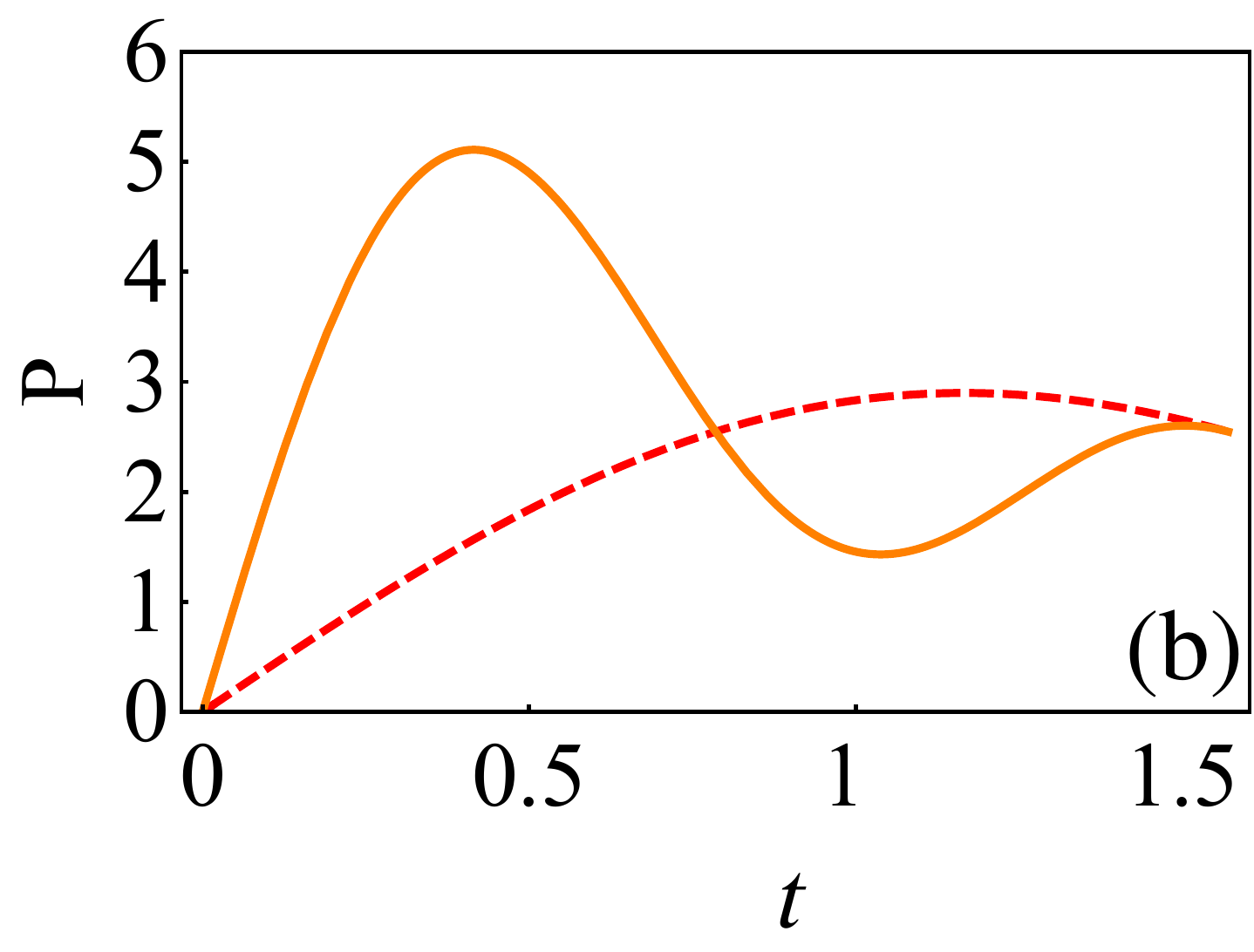}}
\caption{When $D=0$, the evolution of ergotropy ${\zeta }$  and charging power ${P} $ evolve with time ${\Omega t} $ under collective charging and parallel charging
in the Ising model. The collective charging mode (orange solid line), and the parallel charging mode (red dashed line). Herein, ${J=\Omega}$ is held for the
coupling strength and magnetic field.}
\label{f2}
\end{figure}

To probe the influence of parallel and collective charging on ergotropy and power, we first consider the cases without DM-interaction, and Fig. \ref{f2} has depicted the evolution of the ergotropy and power as a function of the time $t$, for $J=D=0$ (parallel charging mode) and $D=0$ and $J\ne 0$ (collective charging mode).
In the collective charging mode, the ergotropy of the battery oscillatorily rises and reaches the peak ${\zeta _{\max}}$, whose value is identical to that in the parallel charging mode at the same time. Notably, although the ergotropy is the same for the two charging during a period interval $[0,t_{\rm min}]$, the powers between them are different in the interval. Specifically, for $t\in [0,\pi/(4\Omega)]$, the power in collective charging is always higher than that in the parallel one; this can be interpreted as the quantum correlation between cells increasing the charging power of QBs in the current scenario.

\begin{figure}[htbp]
\centering
\subfigure{\includegraphics[width=0.5\linewidth]{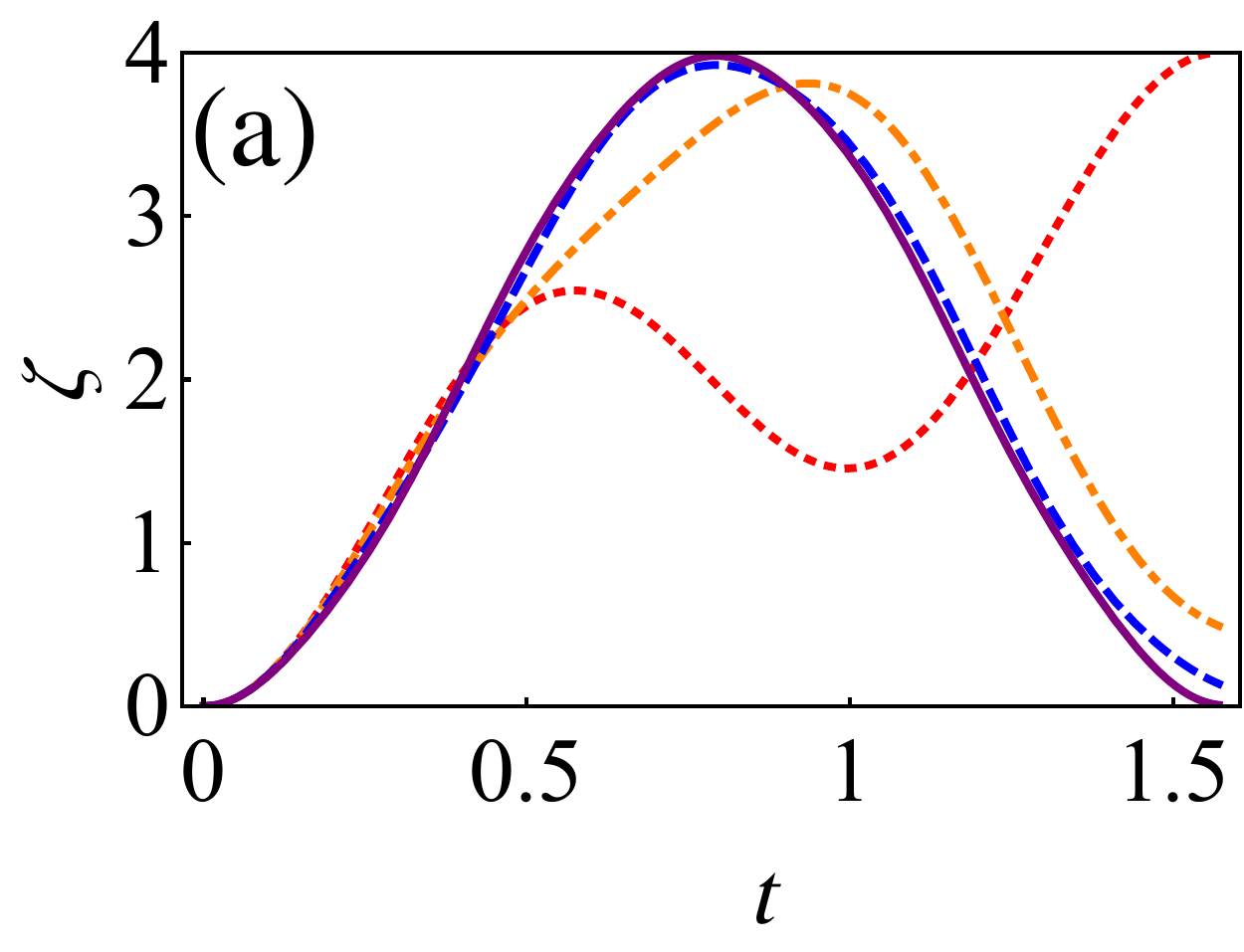}}\subfigure{\includegraphics[width=0.5\linewidth]{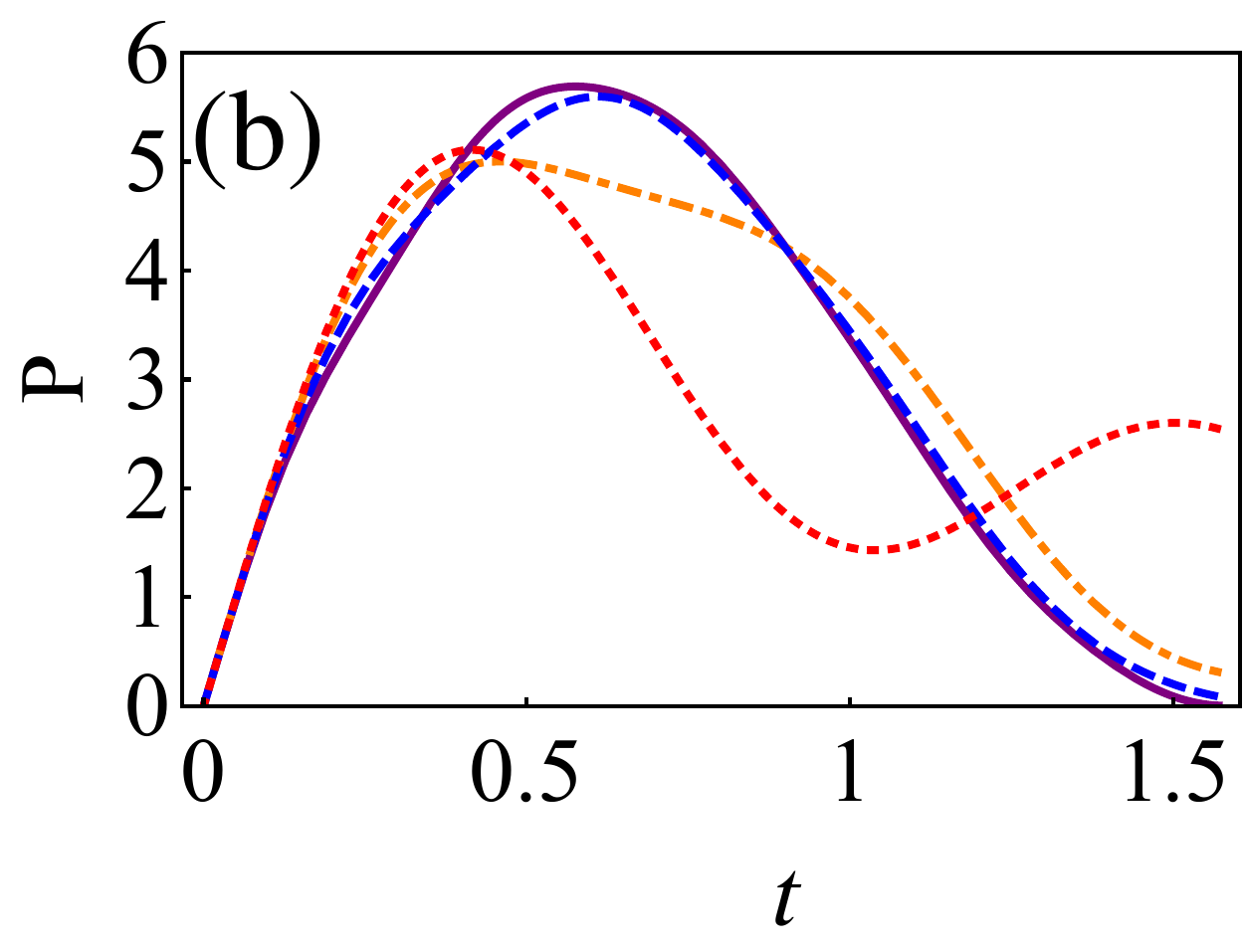}}
\subfigure{\includegraphics[width=8cm]{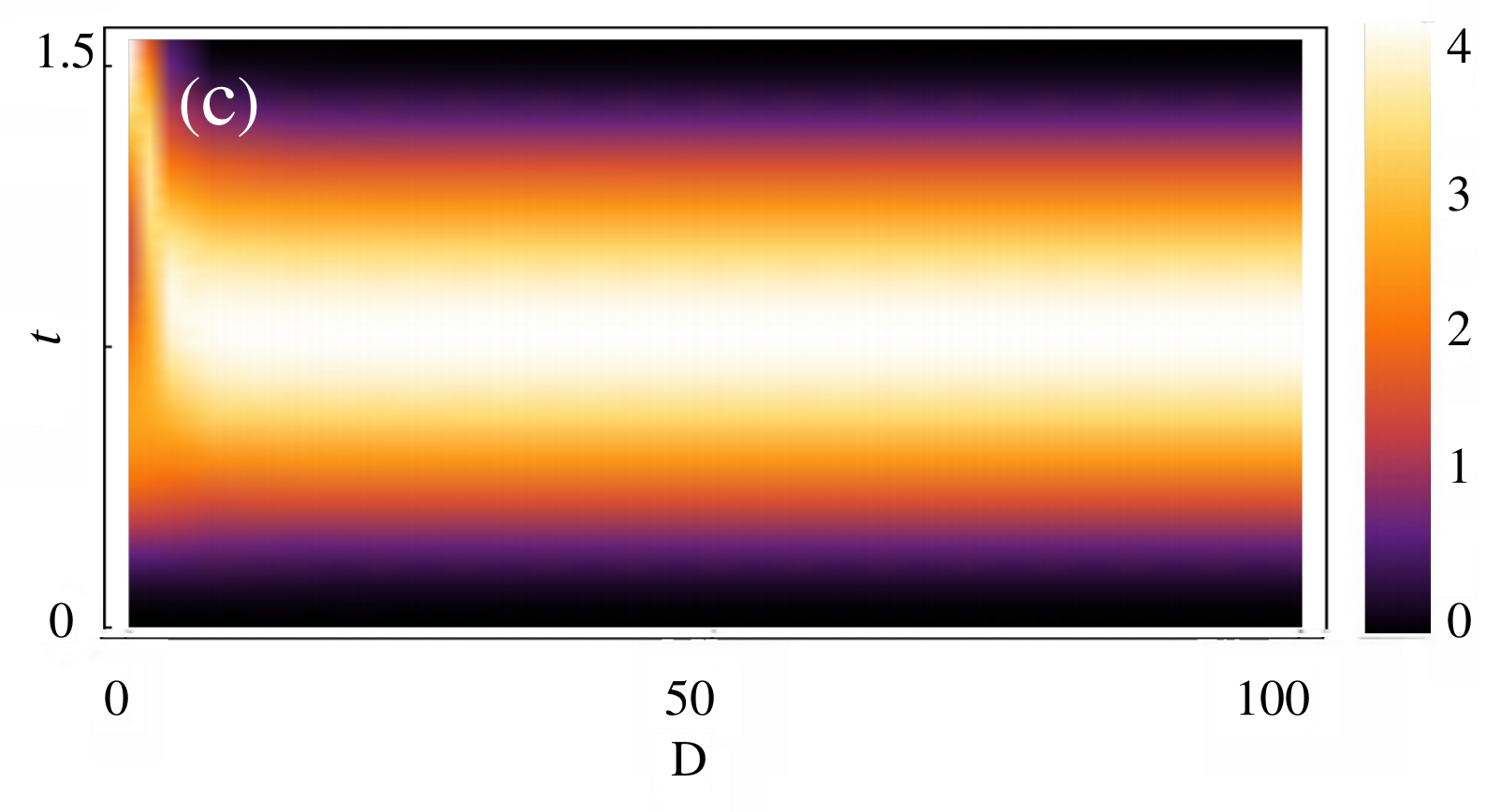}}
\caption{(a)-(b) The ergotropy ${\zeta }$ and charging power ${P} $ evolve with time ${\Omega t} $ in the Ising model. Varying the intensity of DM interactions
in the spin.
The interaction intensities {(in units of $\Omega$)} $D=0$ (the red dotted line),  $D=3$ (orange dot-dashed line), $D=6$ (blue dashed line), $D=9$ (purple solid line) are set
respectively. {(c) The ergotropy versus the time ${\Omega t} $ and the interaction intensity $D$.} Herein, ${J=\Omega}$ is held for the
coupling strength and magnetic field.}
\label{f3}
\end{figure}

We proceed by considering the effect of the {DM} interaction (i.e., ${D}$ $\ne$ 0) on the QB's performance in the Ising model. Fig. \ref{f3} plotted the variation of the battery energy and power over time
for different DM interaction intensities. One can see that the ergotropy can faster reach peak at a moment less
than ${t _{\min}}$ with the stronger DM interaction strength, as shown in Fig. \ref{f3}(a). Additionally, the maximal powers
are generally increased with the growing ${D}$, indicated as Fig. \ref{f3} (b).
With these in mind, we conclude that the DM interaction helps enhance the QB's performance in the Ising model.
{Following Fig. \ref{f3}, one can realize that the ergotropy can still reach its maximum
value (white area in the figure), when the value of $D$ is much larger than $J$. Moreover, it is found that the increase of $D$ has little impact on the ergotropy  during the
charging when $D$ is relatively strong.}

\subsection{{XXZ Model}}
\begin{figure}[htbp]
\centering
\subfigure{\includegraphics[width=0.5\linewidth]{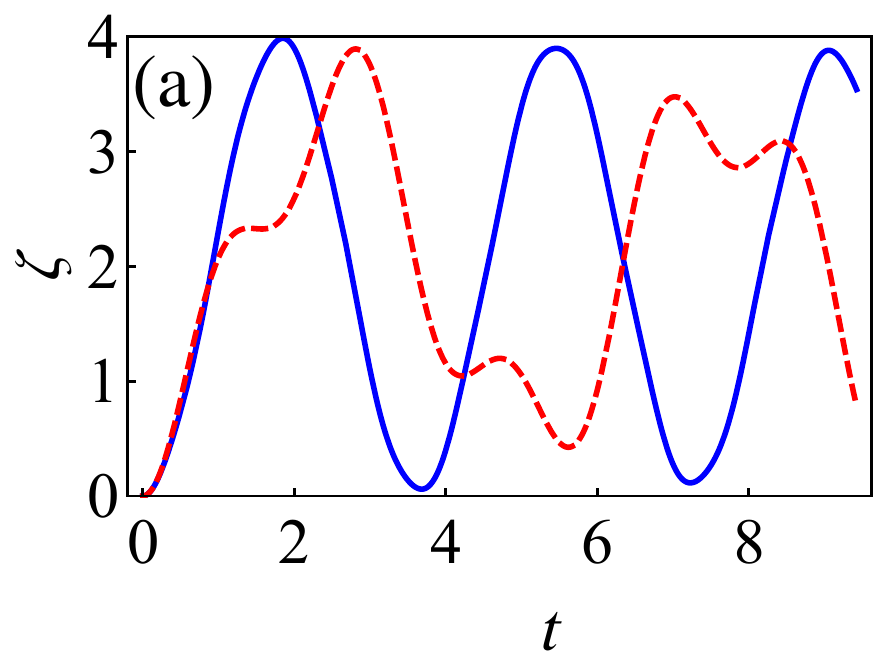}}\subfigure{\includegraphics[width=0.5\linewidth]{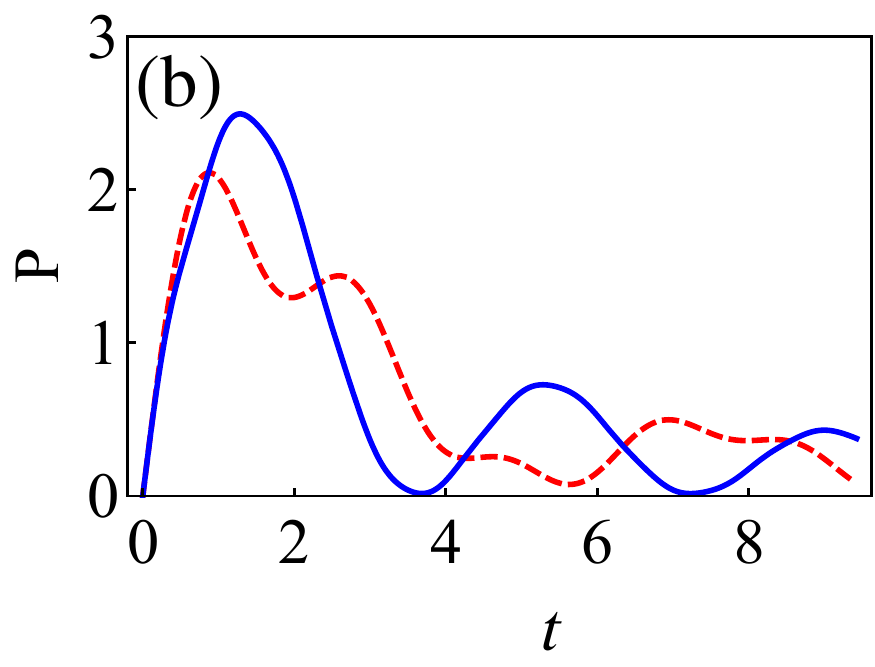}}
\caption{The ergotropy ${\zeta }$  and charging power ${P} $ evolve with time ${\Omega t} $ in the XXZ model. Varying the intensity of DM interactions in the spin.
The interaction intensities {(in units of $\Omega$)} $D=0$ (the dashed red line), and $D=1.7$ (solid blue line). Different timescales have been used
to properly show the maximum position of the related performance. Herein, ${J=\Omega}$ is held for the
coupling strength and magnetic field. And here the anisotropy parameter   ${\Delta = 2}$.}
\label{f4}
\end{figure}

Next, we consider how the effect of {DM} interaction on QB's performance in the {XXZ} model. As mentioned before, the examined system becomes an {XXZ} model with {DM} interaction with respect to ${\gamma =0}$ and ${\Delta \ne 0}$.
From Fig. \ref{f4}(a), we can see that with the progress of charging, the charging
speed of the battery with {DM} interaction becomes faster, and the peak of ergotropy is further increased.
Compared with the case without DM interaction ${D=0}$, the battery with {DM} interaction can achieve a better charging effect in a relatively short time. Moreover, compared with the red line, the blue line maintains a better trend in the long-term
evolution. {That is because the energy between the systems is constantly transferred between the battery and the charger (field) \cite{PhysRevB.99.035421}, causing fluctuations in the peak energy of the battery. DM interaction can facilitate energy transfer between systems, making the fluctuation of the maximal ergotropy relatively small}.
Furthermore, as for the power, Fig. \ref{f4}{(b)} directly indicates that the maximal power of QB with DM interaction is stronger than that without DM interaction. Thereby,
combining these points, the DM interaction can enhance ergotropy and power in the current {XXZ} model, which agrees with the result in the previous subsection.

\begin{figure} [htbp]
\centering
\subfigure{\includegraphics[width=0.51\linewidth]{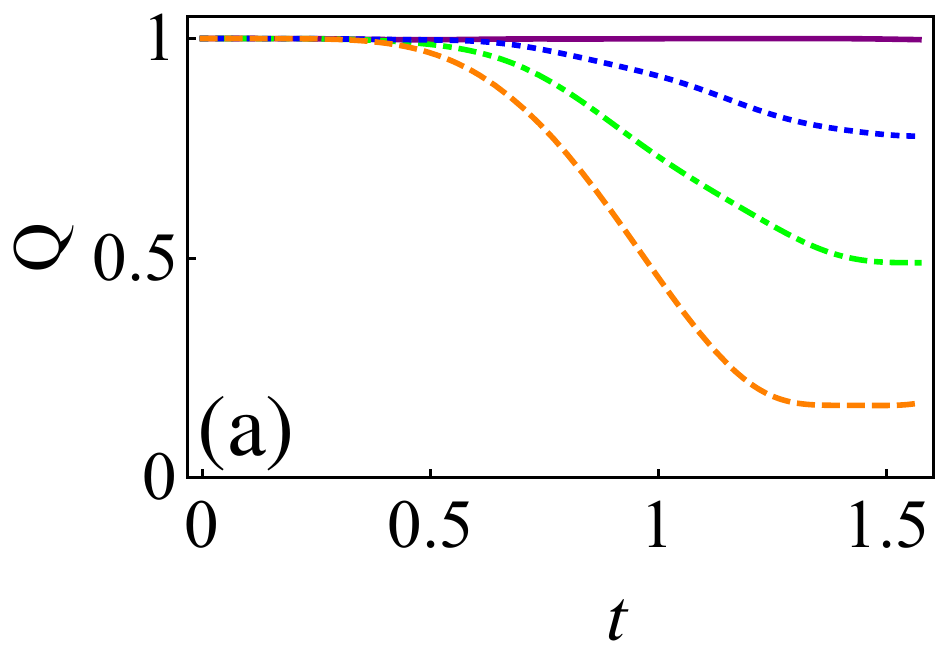}} \subfigure{\includegraphics[width=0.48\linewidth]{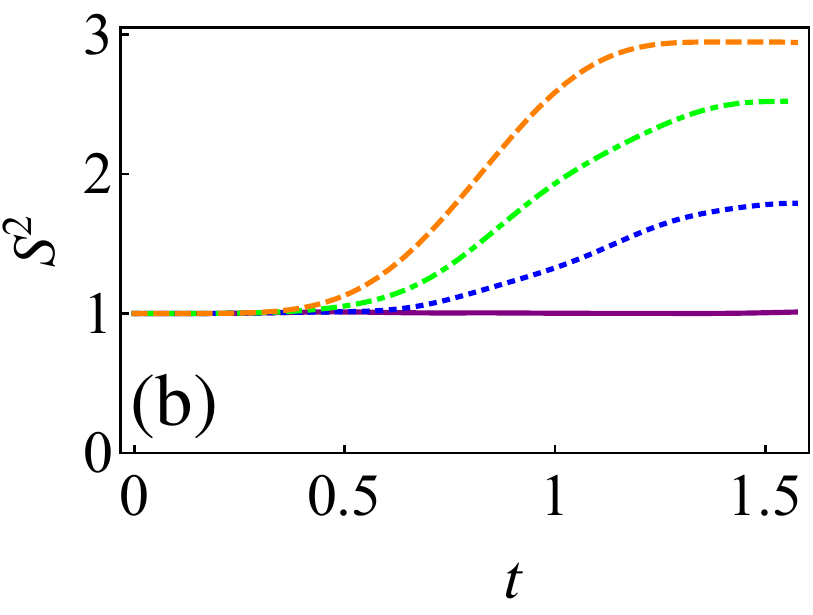}}
\caption{In the XXZ model, the change of first-order coherence with time ${\Omega t}$ under different {DM} interaction intensities and the corresponding change of battery steerability with time ${\Omega t}$
under different ${D}$ values. The DM interaction intensities {(in units of $\Omega$)} $D=0$ (the dashed orange line), $D=0.8$ (the dot-dashed green line), $D=1.2$ (the dotted blue line), and $D=1.7$ (the solid purple line) are set.
Herein, ${J=\Omega}$ is held for the
coupling strength and magnetic field and the anisotropy parameter ${\Delta = 2}$.}
\label{f5}
\end{figure}

In addition, to explain why the DM interaction affects the performance of QB in the above models, we explore the system's quantumness, i.e., quantum first-order coherence and steering.
We plotted the first-order coherence of the system over time in Fig. \ref{f5}(a) and the steerability during this charging
process in Fig. \ref{f5}(b), as for different {DM} interactions. The figure illustrates that larger ${D}$ values correspond to larger first-order coherence throughout the evolution window and weaker steerability between systems. As mentioned previously, the {DM} interaction can enhance ergotropy and power of QBs. Therefore, these suggest that the first-order coherence is in favour of the enhancement of the QB's performance, while the steering between batteries does not positively impact battery energy storage.


\subsection{{XYZ Model}}
\begin{figure}[t]
\centering
\subfigure{\includegraphics[width=0.483\linewidth]{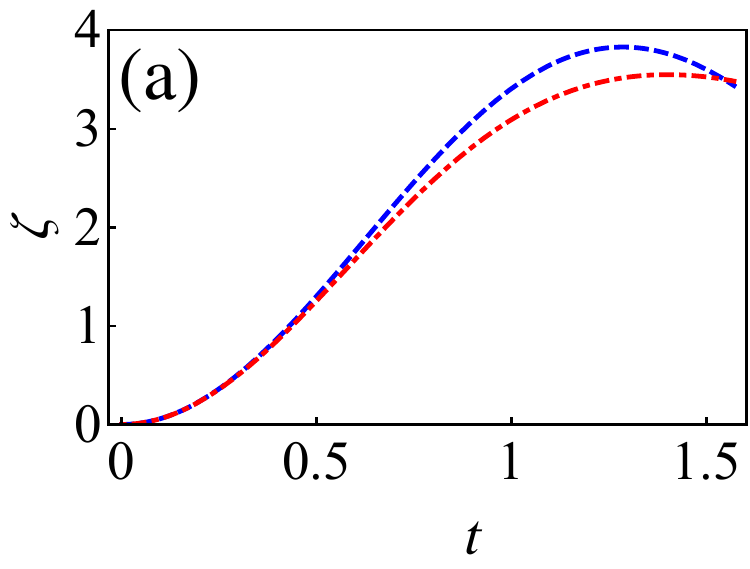}}\subfigure{\includegraphics[width=0.483\linewidth]{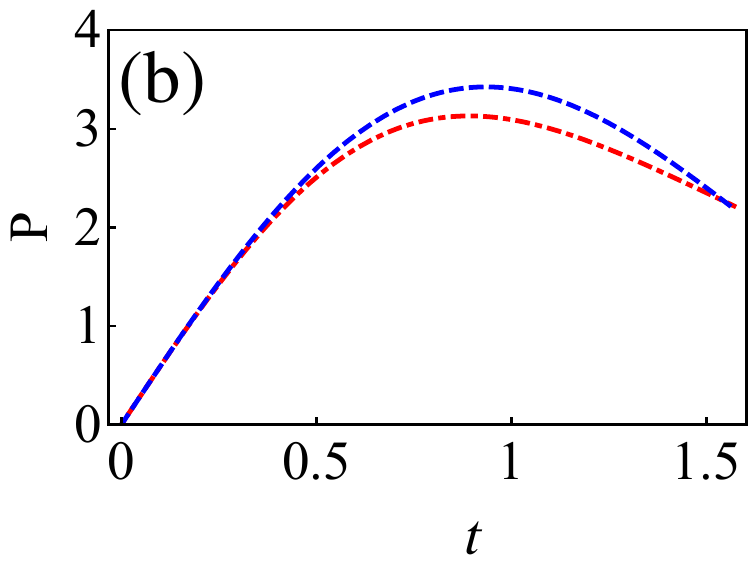}}
\caption{For the XYZ model, time evolution for the ergotropy ${\zeta }$ of the two-cell QB for different values of the anisotropy parameter in the absence of the DM interaction strength.
The anisotropy parameters $\Delta=2.5$ (dashed blue line) and $\Delta=3$ (dotted red line). Herein, ${J=\Omega}$ is set.
}
\label{f6}
\end{figure}

\begin{figure}[t]
\centering
\subfigure{\includegraphics[width=0.483\linewidth]{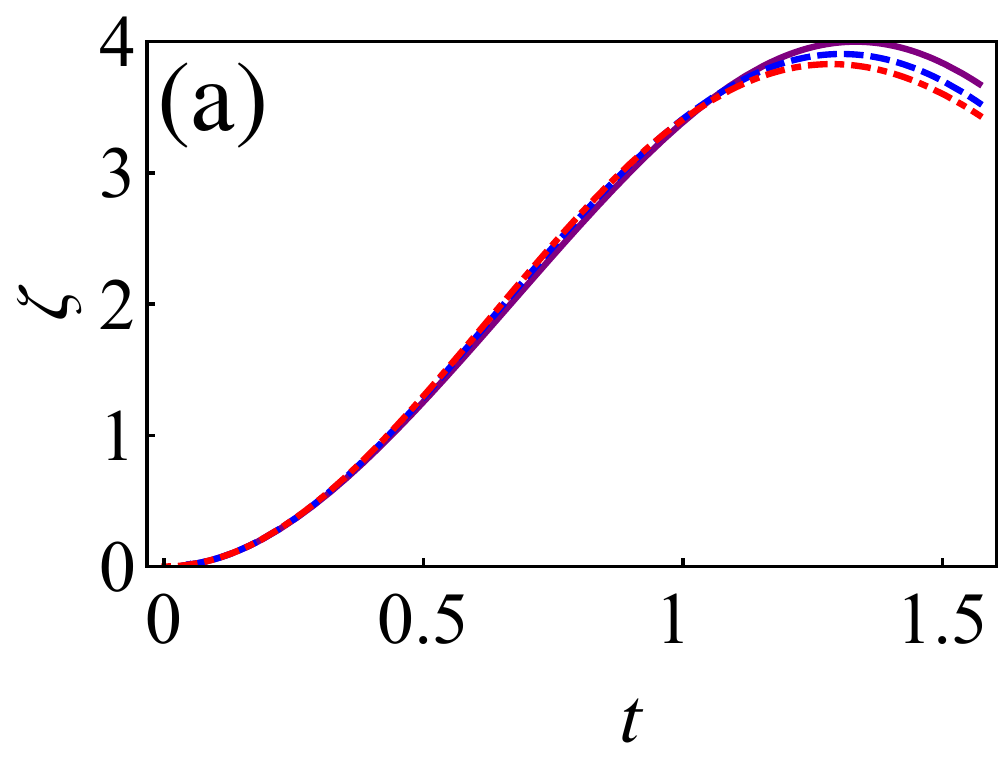}}\subfigure{\includegraphics[width=0.483\linewidth]{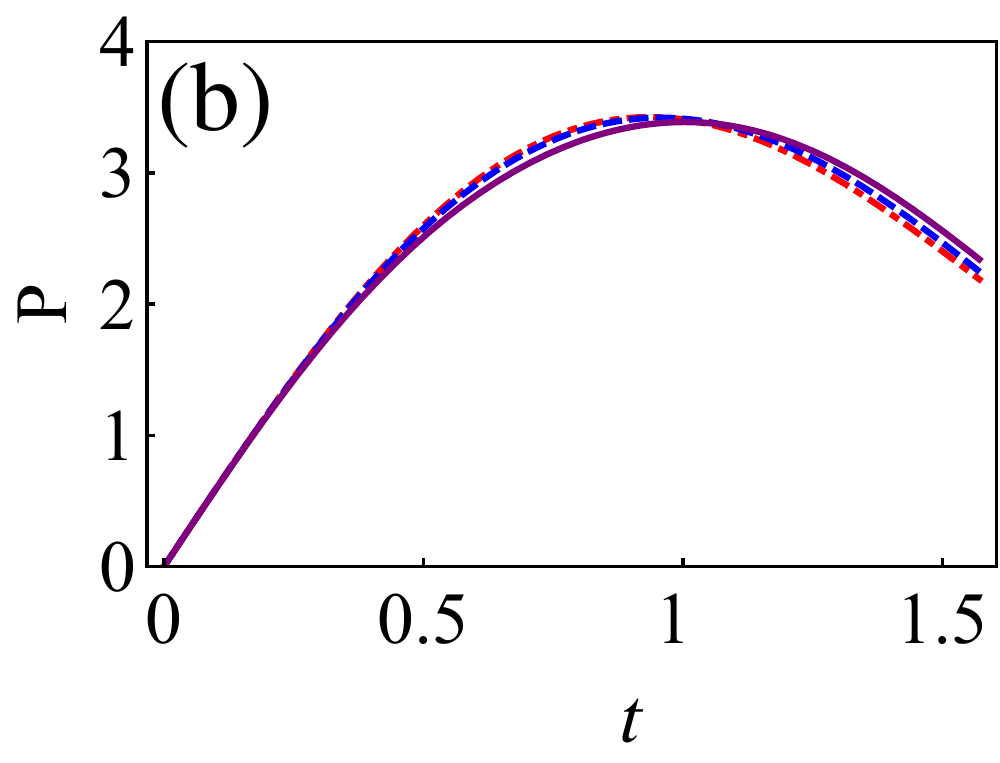}}
\subfigure{\includegraphics[width=0.483\linewidth]{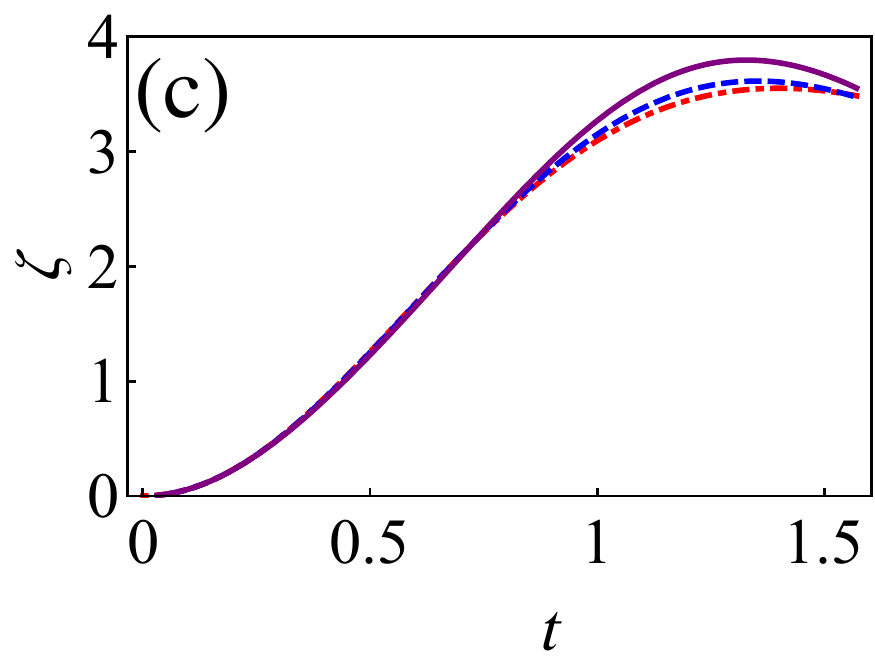}}\subfigure{\includegraphics[width=0.483\linewidth]{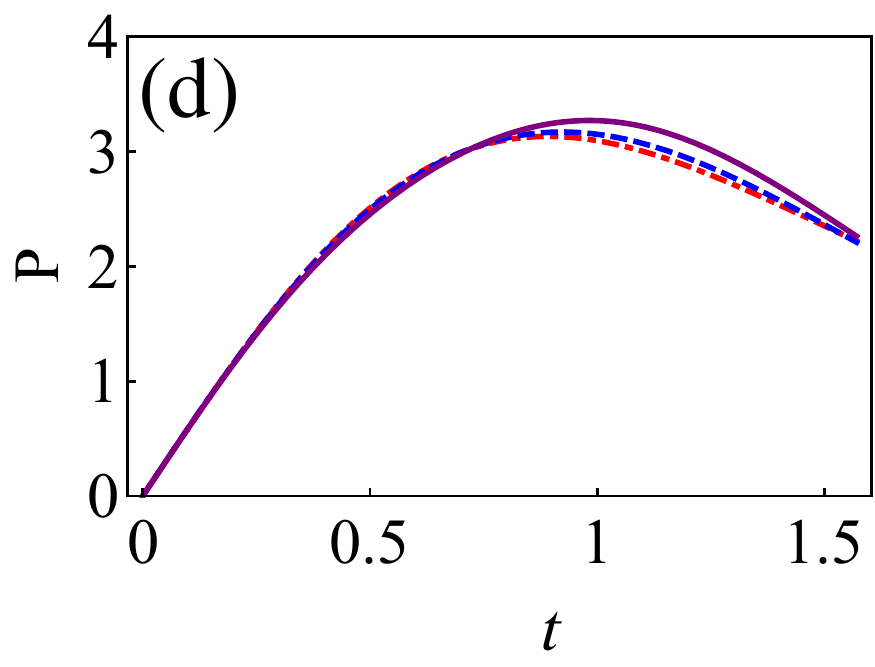}}
\caption{For the XYZ model, the ergotropy ${\zeta }$ and charging power ${P} $ evolution with time when anisotropy parameters ${\Delta= 2.5}$ and ${\Delta=3 }$.
The interaction intensities {(in units of $\Omega$)} $D=0$ (the red dotted line), $D=0.5$ (blue dashed line), $D=1$ (purple solid line). ${J=\Omega}$ is set.}
\label{f7}
\end{figure}

In the above sections, we found that the DM interaction can strengthen the QB's ergotropy and power to some extent.
Next, we consider what role the {DM} interaction will play in the
charging process of {XYZ} model. First,
without the DM-interaction with ${D=0}$, Fig. \ref{f6} describes the ergotropy and power as a function of time $\Omega t$ with respect to ${\Delta=2.5}$ and ${\Delta =3}$ respectively.
From the figure, although the maximums of both ergotropy and power can reach with less time than ${t_{\min}}=\pi/(2\Omega)$, the magnitudes will be smaller than those of the parallel charging.
Conversely, we have plotted the dynamics of ergotropy and power with the DM interaction in Fig. \ref{f7}. It shows that
the ergotropy and power of the battery are greater than those without {DM} interaction (red line in the figure); this implies that the DM interaction can also improve
 the performance of QBs in the current architecture.

\section{Discussions and conclusions}
In summary, we have investigated the model of quantum batteries with one-dimensional Heisenberg spin chains with and without {DM} interactions. Explicitly, we examined and compared
the properties of two types of charging models, i.e., collective charging and parallel charging, in detail, and unveiled the
influence of {DM} interaction on the charging process of the quantum battery. Interestingly, we found that the DM interaction can increase the ergotropy and charging power of the battery in the three spin-chain models, i.e., the Ising, XXZ, and XYZ models. Thus, we claimed that it could improve the charging performance of QBs in the current architectures,
indicating a better charging effect can be obtained during collective charging compared with parallel charging. Meanwhile,
it is found that the first-order coherence between the systems is a useful resource for battery charging, while the steering between the batteries is not conducive to the storage of battery energy.
{Moreover, it is argued that the current investigation, in principle, might be generalized to $N$-cells QBs.
Theoretically, the number of cells directly affects how much energy a battery can store, and generally, more cells are beneficial to store more energy in the QBs  \cite{z1,z54,z55}.
Thus, we conjecture that the DM interaction can further enhance the battery performance as the number of cells increases and possibly induce
quantum phase transitions and critical behaviours   \cite{z56}.}
Overall, our investigation sheds light on the positive role of the DM interaction in quantum batteries under the spin-chain models and is beneficial for understanding the properties of
practical spin-based quantum batteries.

\begin{acknowledgements}
This work was supported by the National Natural Science Foundation of China (Grant Nos. 12075001 and 61601002), and
Anhui Provincial Key Research and Development Plan (Grant No. 2022b13020004).
\end{acknowledgements}

\begin{references}
\bibitem{z1}F. Campaioli, S. Gherardini, J. Q. Quach, M. Polini, G. M. Andolina, 
\textit{arXiv}:2308.02277

\bibitem {z2} L. Wang, S.-Q. Liu, F.-L. Wu, H. Fan, S.-Y. Liu, 
\textit{Phys. Rev. A {\bf 2023}, 108}, 062402.

\bibitem {z3} W.-L. Song, H.-B. Liu, B. Zhou, W.-L. Yang, J.-H. An, 
\textit{Phys. Rev. Lett. {\bf 2024}, 132}, 090401.

\bibitem {z4} B. \ifmmode \mbox{\c{C}}\else \c{C}\fi{}akmak, 
\textit{Phys. Rev. E {\bf 2020}, 102}, 042111.

\bibitem {z5} S. Juli\`a-Farr\'e, T. Salamon, A. Riera, M. Bera, M. Lewenstein, 
\textit{Phys. Rev. Res. {\bf 2020}, 2}, 023113.

\bibitem {z6} L. Fusco, M. Paternostro, G. D. Chiara, 
\textit{Phys. Rev. E {\bf 2016}, 94}, 052122.

\bibitem {z7} F. Barra, 
\textit{Phys. Rev. Lett. {\bf 2019}, 122}, 210601.

\bibitem {z8} D. Farina, G. M. Andolina, A. Mari, M. Polini, V. Giovannetti, 
\textit{Phys. Rev. B {\bf 2019}, 99}, 035421.

\bibitem {z9} J. Q. Quach, Wi. J. Munro, 
\textit{Phys. Rev. Appl. {\bf 2020}, 14}, 024092.
.
\bibitem {z10} R. Alicki and M. Fannes, 
\textit{Phys. Rev. E {\bf 2013}, 87}, 042123.

\bibitem {z11} T. F. F. Santos, Y. V. D. Almeida, M. F. Santos, 
\textit{Phys. Rev. A {\bf 2023}, 107}, 032203.

\bibitem {z12} M.-L. Song, L.-J. Li, X.-K. Song, L. Ye, D. Wang, 
\textit{Phys. Rev. E {\bf 2022}, 106}, 054107.

\bibitem {z13} M.-L. Song, X.-K. Song, L. Ye, and D. Wang, 
\textit{Phys. Rev. E {\bf 2024}, 109}, 064103.

\bibitem {z14} S. Gherardini, F. Campaioli, F. Caruso, F. C. Binder, 
\textit{Phys. Rev. Res. {\bf 2020}, 2}, 013095.

\bibitem {z15} F. Binder, S. Vinjanampathy, K. Modi, J. Goold, 
\textit{Phys. Rev. E {\bf 2015}, 91}, 032119.

\bibitem {z16} K. V. Hovhannisyan, M. Perarnau-Llobet, M. Huber, A. Ac\'{\i}n, 
\textit{Phys. Rev. Lett. {\bf 2013}, 111}, 240401.

\bibitem {z17} F. H. Kamin, F. T. Tabesh, S. Salimi, Alan C. Santos, 
\textit{Phys. Rev. E {\bf 2020}, 102}, 052109.

\bibitem {z18} F.-Q. Dou, Y.-Q. Lu, Y.-J. Wang, J.-A. Sun, 
\textit{Phys. Rev. B {\bf 2022}, 105}, 115405.

\bibitem {z19} F. C. Binder, S. Vinjanampathy, K. Modi, J. Goold, 
\textit{New Journal of Physics {\bf 2015}, 17}, 075015.

\bibitem {z20} W. J. Zhang, S. Y. Wang, C. F. Wu, G. C. Wang, 
\textit{Phys. Rev. E {\bf 2023}, 107}, 054125.

\bibitem {z21} S. S. Seidov, S. I. Mukhin, 
\textit{Phys. Rev. A {\bf 2024}, 109}, 022210.

\bibitem {z22} K. Sen, U. Sen, 
\textit{Phys. Rev. A {\bf 2021}, 104}, L030402.

\bibitem {z23} A. C. Santos, 
\textit{Phys. Rev. E {\bf 2021}, 103}, 042118.

\bibitem {z24} Y.-Y. Zhang, T.-R. Yang, L. Fu, X.G. Wang, 
\textit{Phys. Rev. E {\bf 2019}, 99}, 052106.

\bibitem {z25} F.-Q. Dou, H. Zhou, J.-A. Sun, 
\textit{Phys. Rev. A {\bf 2022}, 106}, 032212.

\bibitem {z26} L. Gao, C. Cheng, W.-B. He, R. Mondaini, X.W. Guan, H.-Q. Lin, 
\textit{Phys. Rev. Res. {\bf 2022}, 4}, 043150.

\bibitem {z27} G. M. Andolina, M. Keck, A. Mari, V. Giovannetti, M. Polini, 
\textit{Phys. Rev. B {\bf 2019}, 99}, 205437.

\bibitem {z28} W. Chang, T.-R. Yang, H. Dong, L.B. Fu, X. G. Wang, Y.-Y. Zhang, 
\textit{New Journal of Physics {\bf 2021}, 23}, 103026.

\bibitem {z29} L. Peng, W.-B. He, S. Chesi, H.-Q. Lin, X.-W. Guan, 
\textit{Phys. Rev. A {\bf 2021}, 103}, 052220.

\bibitem {z30} F. Pirmoradian, K. M\o{}lmer, 
\textit{Phys. Rev. A {\bf 2019}, 100}, 043833.

\bibitem {z31} J.-Y. Gyhm, D. \ifmmode \check{S}\else \v{S}\fi{}afr\'anek, D. Rosa, 
\textit{Phys. Rev. Lett. {\bf 2022}, 128}, 140501.

\bibitem {z32} J. Kim, J. Murugan, J. Olle, D. Rosa, 
\textit{Phys. Rev. A {\bf 2022}, 105}, L010201.

\bibitem {z33} F. Campaioli, F. A. Pollock, F. C. Binder, L. C\'eleri,  J. Goold, S. Vinjanampathy, K. Modi, 
\textit{Phys. Rev. Lett. {\bf 2017}, 118}, 150601.

\bibitem {z34} T. P. Le, J. Levinsen, K. Modi, M. M. Parish, F. A. Pollock, 
\textit{Phys. Rev. A {\bf 2018}, 97}, 022106.

\bibitem {z35} H.-L. Shi, S. Ding, Q.-K. Wan, X.-H. Wang, W.-L. Yang, 
\textit{Phys. Rev. Lett. {\bf 2022}, 129}, 130602.

\bibitem {z36} I. Dzyaloshinsky, 
\textit{Journal of Physics and Chemistry of Solids {\bf 1958}, 4}, 241-255.

\bibitem {z37} T. Moriya, 
\textit{Phys. Rev. {\bf 1960}, 120}, 91--98.

\bibitem {z38} A. C. Santos, A. Saguia, M. S. Sarandy, 
\textit{Phys. Rev. E {\bf 2020}, 101}, 062114.

\bibitem {z39} A. E. Allahverdyan, R. Balian, Th. M. Nieuwenhuizen, 
\textit{Europhysics Letters {\bf 2004}, 67}, 565.

\bibitem {z40} A. C. Santos, B. \ifmmode \mbox{\c{C}}\else \c{C}\fi{}akmak, S. Campbell, N. T. Zinner, 
\textit{Phys. Rev. E {\bf 2019}, 100}, 032107.

\bibitem {z41} R. Uola, A. C. S. Costa, H. C. Nguyen, O. G\"uhne, 
\textit{Rev. Mod. Phys. {\bf 2020}, 92}, 015001.

\bibitem {z42} A. F. Abouraddy, M. B. Nasr, B. E. A. Saleh, A. V. Sergienko, M. C. Teich, 
\textit{Phys. Rev. A {\bf 2021}, 63}, 063803.

\bibitem {z43} P. S. Burada, G. Schmid, 
\textit{Phys. Rev. E {\bf 2010}, 82}, 051128.

\bibitem {z44} N. Brunner, D. Cavalcanti, S. Pironio, V. Scarani, S. Wehner, 
\textit{Rev. Mod. Phys. {\bf 2014}, 86}, 419--478.

\bibitem {z45} M. T. Quintino, T. V\'ertesi, Da. Cavalcanti, R. Augusiak, M. Demianowicz, A. Ac\'{\i}n , N. Brunner, 
\textit{Phys. Rev. A {\bf 2015}, 92}, 032107.

\bibitem {z46} H. M. Wiseman, S. J. Jones, A. C. Doherty, 
\textit{Phys. Rev. Lett. {\bf 2007}, 98}, 140402.

\bibitem {z47} E. G. Cavalcanti, S. J. Jones, H. M. Wiseman, M. D. Reid, 
\textit{Phys. Rev. A {\bf 2009}, 80}, 032112.

\bibitem {z48} A. C. S. Costa , R. M. Angelo, 
\textit{Phys. Rev. A {\bf 2016}, 93}, 020103.

\bibitem {z49} K. H. Kagalwala, G. D. Giuseppe, A. F. Abouraddy, B. E. A. Saleh, 
\textit{Nature Photonics {\bf 2012}, 7}, 72-78.

\bibitem {z50}  J. K. Kalaga, W. Leo\ifmmode \acute{n}\else \'{n}\fi{}ski,  J. Pe\ifmmode \check{r}\else \v{r}\fi{}ina, 
\textit{Phys. Rev. A {\bf 2018}, 97}, 042110.

\bibitem {z51} D.-D. Dong, X.-K. Song, X.-G. Fan, L. Ye, D. Wang, 
\textit{Phys. Rev. A {\bf 2023}, 107}, 052403.

\bibitem {z52} X.-G. Fan, W.-Y. Sun, Z.-Y. Ding, F. Ming, H. Yang, D. Wang, L. Ye, 
\textit{New Journal of Physics {\bf 2019}, 21}, 093053.

\bibitem {z53} L. Mandel, E. Wolf, Optical Coherence and Quantum Optics,
\textit{(Cambridge University Press {\bf 1995})}.

\bibitem {z54} D. Ferraro, M. Campisi, G. M. Andolina, V. Pellegrini, M. Polini, 
\textit{Phys. Rev. Lett. {\bf 2018}, 120}, 117702.

\bibitem {z55} W.-X. Guo, F.-M. Yang, F.-Q. Dou, 
\textit{Phys. Rev. A {\bf 2024}, 109}, 032201.

\bibitem {z56} F.-W. Ma, S.-X. Liu, X.-M. Kong, 
\textit{Phys. Rev. A {\bf 2011}, 84}, 052109.

\end{references}

\end{document}